\documentclass{PoS}

\usepackage{wrapfig,bm}

\title{$\bm{|V_{cb}|}$ using lattice QCD}

\ShortTitle{$|V_{cb}|$ using lattice QCD}

\author{\speaker{Matthew Wingate}\\
        Department of Applied Mathematics and Theoretical Physics,
        University of Cambridge,
        Cambridge CB3 0WA, United Kingdom\\
        E-mail: \email{M.Wingate@damtp.cam.ac.uk}}

\abstract{Lattice QCD calculations of hadronic matrix elements allow
  one to draw inferences about quark flavor interactions from
  measurements of hadron decays.  Within the context of the Standard
  Model, the magnitude of the charm-bottom quark coupling $V_{cb}$ can
  be determined from semileptonic decays such as $B\to
  D^{(*)}\ell\nu$.  This brief review summarizes the present status
  and short-term outlook for determining $|V_{cb}|$ using lattice QCD.
}

\FullConference{9th International Workshop on the CKM Unitarity Triangle\\
         28 November - 3 December 2016\\
         Tata Institute for Fundamental Research (TIFR), Mumbai, India}

\begin{document}

\section{Introduction}
\label{sec:intro}

\begin{wrapfigure}[16]{R}{0.5\textwidth}
  \centering
  \vspace{-4truemm}
  \includegraphics[width=0.5\textwidth]{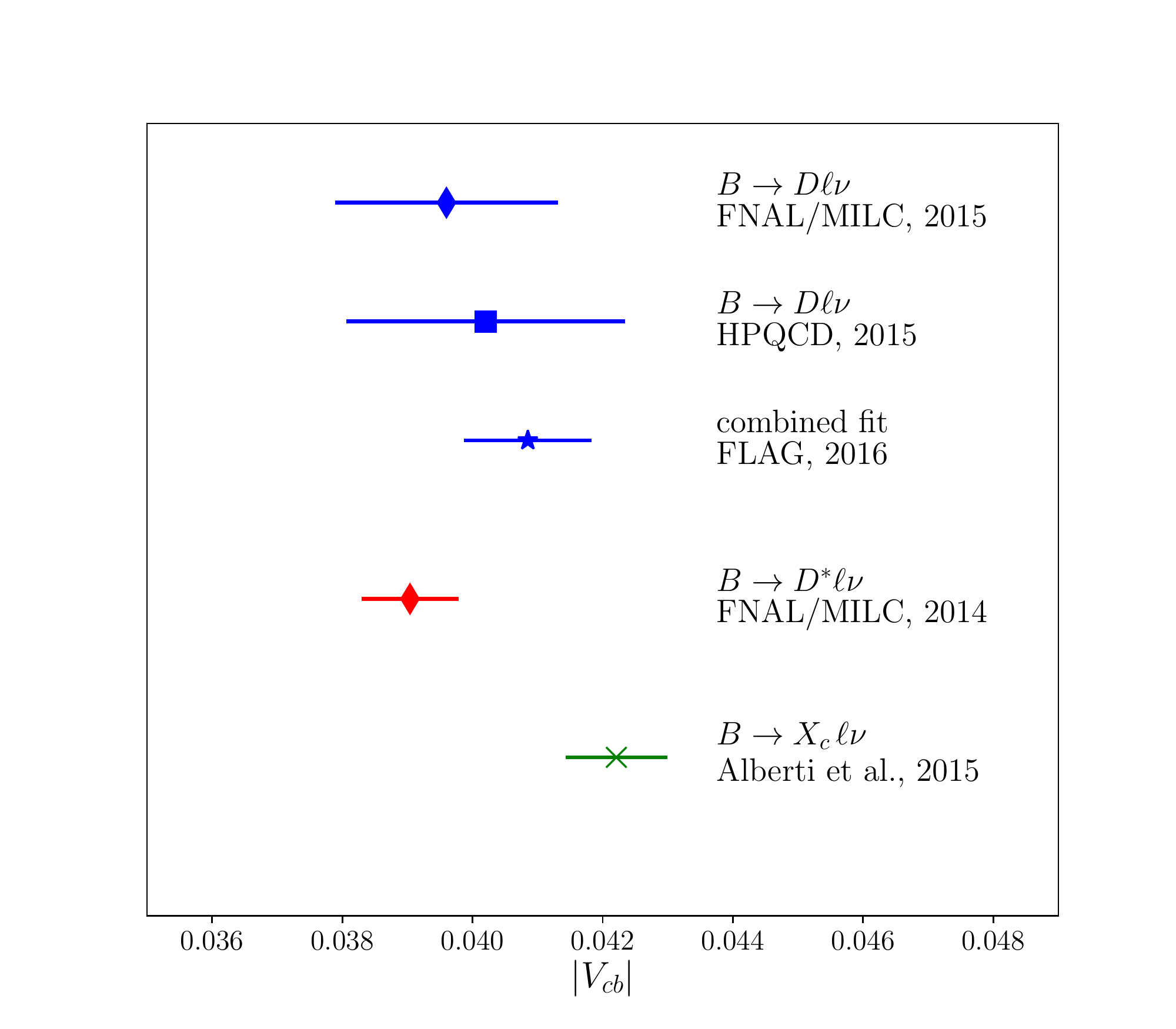}
  \vspace{-5truemm}
  \caption{\label{fig:compare_Vcb}$|V_{cb}|$ from different inputs.}
\end{wrapfigure}

At this workshop the CKM matrix element $V_{cb}$ needs no introduction.
Its present estimates are summarized in Fig.~\ref{fig:compare_Vcb}.
The most precise determination of $|V_{cb}|$ using an exclusive decay
mode comes from combining experimental results for $B\to D^*\ell\nu$
with the relevant form factor at zero recoil 
\cite{Amhis:2012bh,Bailey:2014tva}.\footnote{Note a new 
$B\to D^*\ell\nu$ result, $(37.4\pm1.3)\times 10^{-3}$, from Belle 
recently appeared \cite{Abdesselam:2017kjf}.}
With similar precision, one can infer $|V_{cb}|$ from inclusive
semileptonic $b\to c$ decays using an operator product expansion
\cite{Bevan:2014iga,Alberti:2014yda}.  These two values disagree at the
$3\sigma$ level.

The determination of $|V_{cb}|$ from $B\to D\ell\nu$ decay has been less
precise due to larger experimental uncertainties.
Figure~\ref{fig:compare_Vcb} shows the
published results from two collaborations
\cite{Bailey:2015rga,Na:2015kha}. The FLAG combined fit of the lattice
results \cite{Aoki:2016frl} including new experimental results
\cite{Glattauer:2015teq} is also shown.  The fact that the $B\to
D\ell\nu$ $|V_{cb}|$ determinations lie between those from $B\to D^*\ell\nu$
and $B\to X_c\ell\nu$ implies that the explanation for the discrepancy
is not due to new physics manifesting itself as a new right-handed
interaction. Such an effect would cause the $B\to D\ell\nu$
determination to be the outlier, since only the vector current
contributes to this decay.

Below I discuss a few more details about published work on this
topic, and I review the status of calculations in progress.  First I
wish to give a brief survey of some lattice QCD details.

\section{Survey of methods}
\label{sec:methods}

\begin{figure}[t]
\centering
\includegraphics[width=0.42\textwidth]{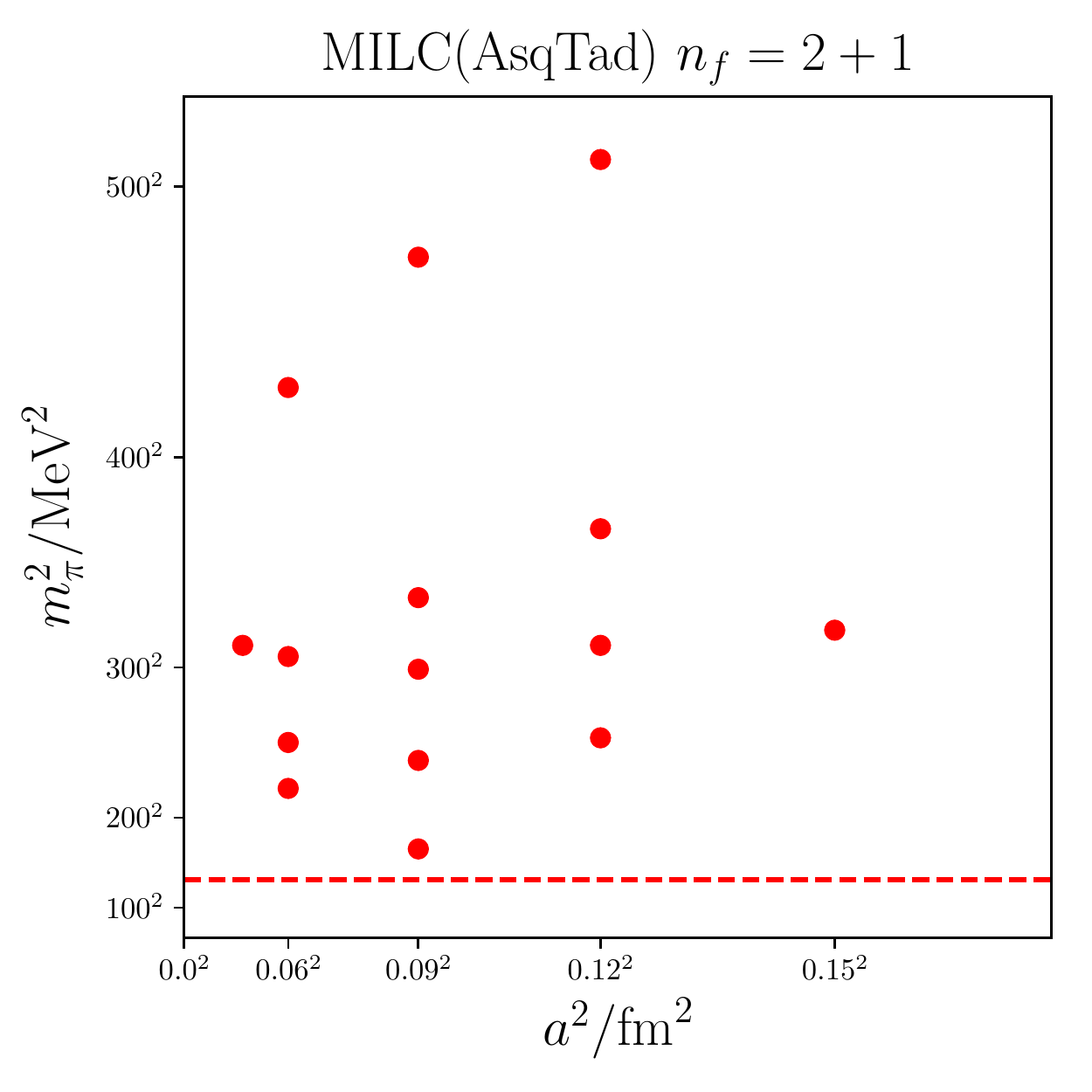}
\includegraphics[width=0.42\textwidth]{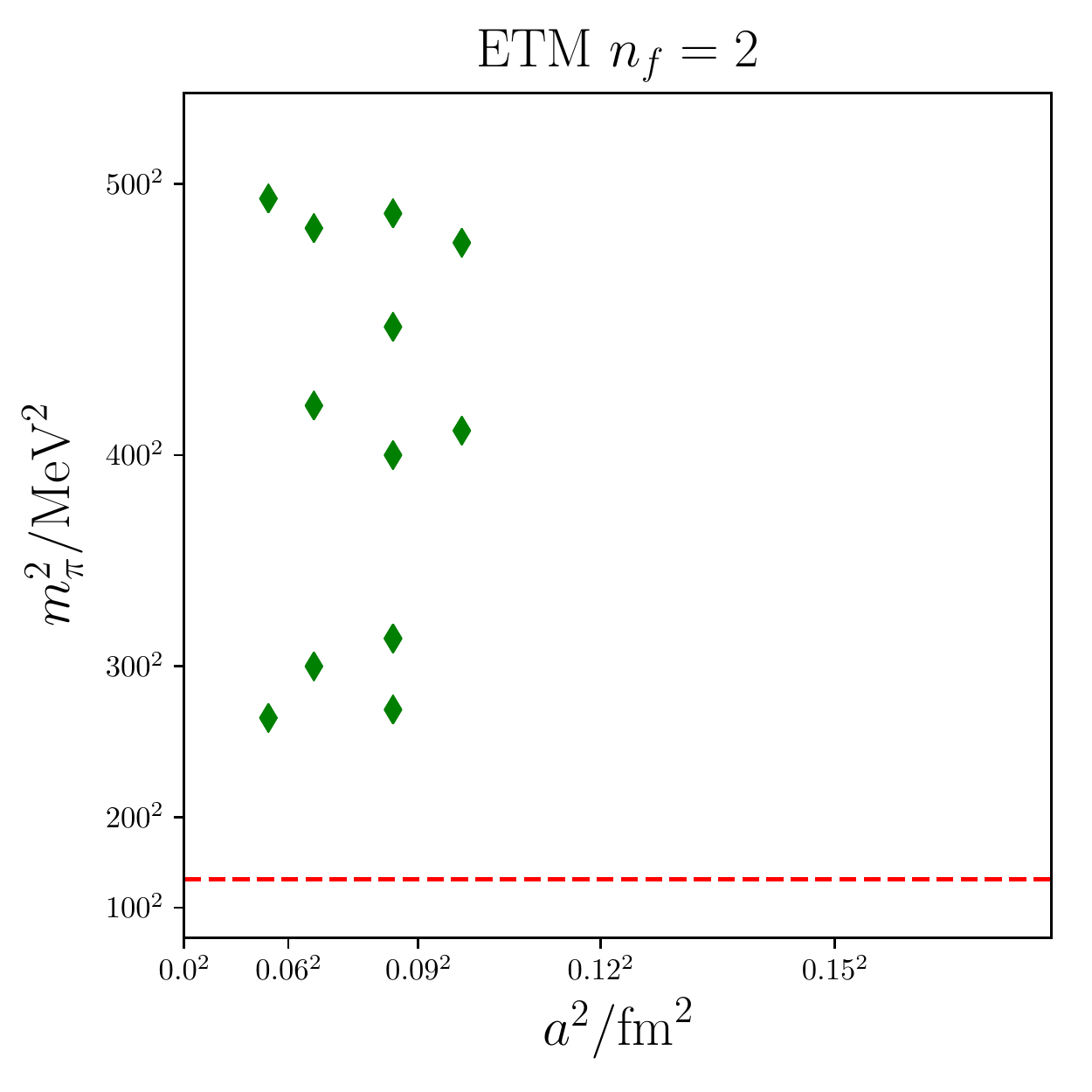}
\includegraphics[width=0.42\textwidth]{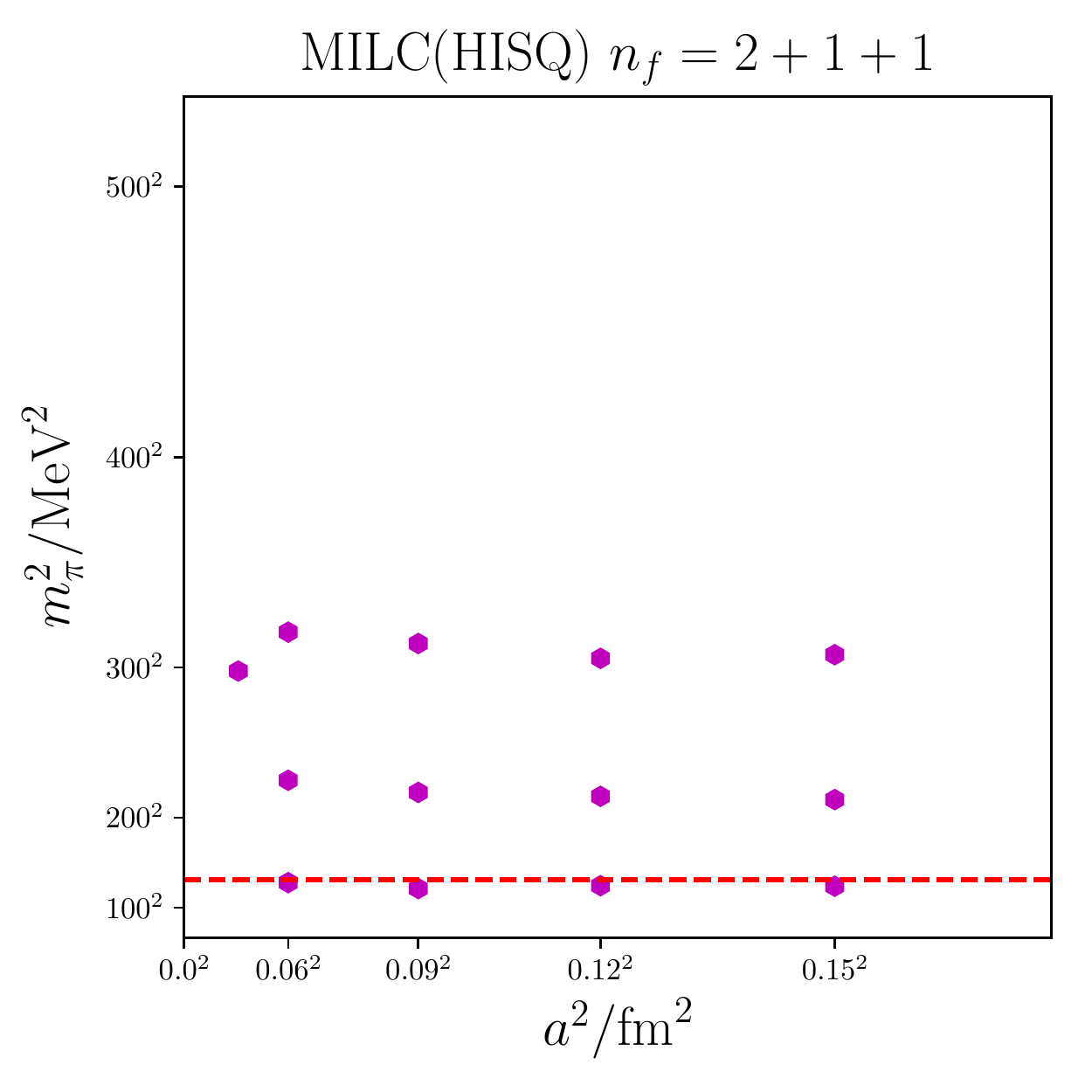}
\includegraphics[width=0.42\textwidth]{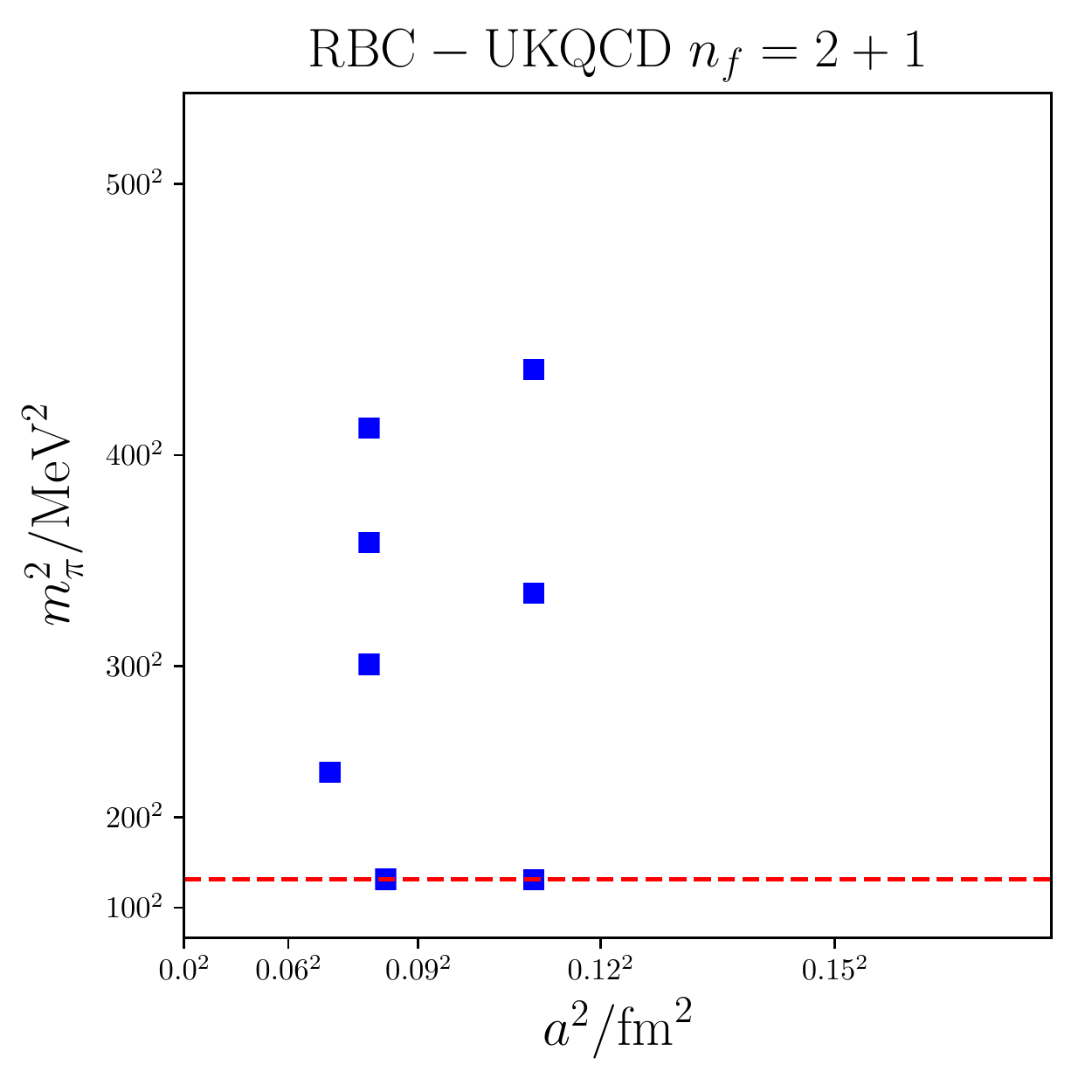}
\caption{\label{fig:ensembles}Plots showing the pion masses and lattice
  spacings for unquenched gauge field configurations with $n_f$ flavours of
  sea quarks. Each plot corresponds to a separate ensemble of configurations,
  differing most significantly in the choice of fermion discretization (see
  text). The red dashed line corresponds to the physical pion mass.}
\end{figure}

The choice of discretization of the Dirac Lagrangian (or action) is a
crucial decision one makes when carrying out a lattice QCD (LQCD)
calculation.  Figure~\ref{fig:ensembles} summarizes (in part) the
unquenched lattice configurations relevant for recent and present
calculations of $|V_{cb}|$; each graph corresponds to a different
lattice action.  The plots show the pion mass, a proxy for the light
quark mass input into the calculation, against the lattice spacing
inferred by requiring some dimensionless lattice output be equal to
its physical value.  Both quantities enter the plot squared, making
extrapolation to the physical limit roughly linear.  For $B$
decays, effects of discretization and unphysical quark mass have been
the most sensitive to get under control.  Of course one needs to
ensure other systematic errors, such as finite volume effects, are
also quantified.

The LQCD results used in Fig.~\ref{fig:compare_Vcb} were all obtained
using improved versions of the staggered quark action, using the MILC
AsqTad configurations.  Staggered quarks are computationally
inexpensive and can be improved to have small discretization errors;
however, an additional assumption is required to include a number of
sea quarks which is not a multiple of 4.  There is a body of
literature discussing this approach, and the empirical evidence
supports its soundness.

The ETM collaboration use twisted mass quarks, an improved variant of
Wilson's action.  The RBC-UKQCD collaboration use domain wall
quarks.  The plots in Fig.~\ref{fig:ensembles} hint at some of the
advantages and disadvantages of the different formulations.  The (relative)
efficiency of staggered fermions allows for calculations to be done
with many different sets of input parameters. The high level
of improvement possible with staggered actions, first the AsqTad variant 
then HISQ, mean that discretization errors are greatly reduced at
larger values of the lattice spacing.  Twisted mass fermions have
sizable discretization errors, so calculations with smaller values of $a$
have been necessary.  Domain wall quarks are 
computationally expensive; however, there is strong motivation to
invest heavily in this approach since they have the continuum-like 
chiral and flavor symmetries.

Heavy quarks on the lattice provide another challenge. The energy
scale at which discretization errors typically become large is given
by the inverse lattice spacing.  The charm quark mass is at or below
this scale, and the bottom quark mass is significantly
higher. Effective field theories have been used to make progress,
especially for $b$ quark physics.  Until recently, the Fermilab
approach to heavy quarks has been only one used to pursue high precision
lattice QCD calculations of $b\to c$ matrix elements.  With the increase
in computational power and more highly improved actions, other approaches
are now able to provide independent checks of Fermilab/MILC results. 

On the MILC AsqTad lattices, the Fermilab/MILC collaboration
constructed correlation functions using the same action for the light
valence quarks and the Fermilab approach for the $c$ and $b$ quarks.
The HPQCD collaboration used the more highly improved, HISQ version
of the staggered action for the valence quarks (both light and charm),
using nonrelativistic QCD (NRQCD) for the $b$ quark.  A subset of the
MILC AsqTad configurations were used by HPQCD, while Fermilab/MILC used
the whole set.  Both groups have used their data to refine the
Standard Model prediction for $R(D)$ the ratio of $B\to D\tau\nu$ to
$B\to D\ell\nu$ branching fractions (where $\ell = e, \mu$)
\cite{Bailey:2012jg,Bailey:2015rga,Na:2015kha}.  Fermilab/MILC have
also used their form factors to estimate the neutral $B$ meson
fragmentation functions, of use in extractions of the rate for
$B_s^0\to \mu^+\mu^-$ \cite{Bailey:2012rr}.

The Paris group have used the ETM twisted-mass configurations
(Fig.~\ref{fig:ensembles}) to calculate the $B_s\to D_s\ell\nu$ form factors
near zero recoil \cite{Atoui:2013zza}.  They 
interpolate a well-chosen ratio between a known result in the static limit and
lattice calculations performed with charm-like masses for the $b$.  Their
quoted results, e.g.\ for the form factor extrapolated to zero recoil, are
consistent with those obtained on MILC lattices, albeit with much larger
uncertainties.

Baryonic decays also play a role in over-constraining $|V_{cb}|$.
LHCb has measured the ratio of branching fractions
$\mathcal{B}(\Lambda_b \to
p\ell\nu)/\mathcal{B}(\Lambda_b\to\Lambda_c\ell\nu)$
\cite{Aaij:2015bfa} which, when combined with LQCD determinations of
the corresponding form factors \cite{Detmold:2015aaa}, constrains
$|V_{ub}/V_{cb}|$ to be $0.083(4)_{\mathrm{stat}}(4)_{\mathrm{sys}}$.
Combining this with $|V_{ub}|$, as determined from $B\to\pi\ell\nu$
decay \cite{Aoki:2016frl} gives $|V_{cb}|=0.044(3)$, in somewhat
better agreement with the higher values plotted in
Fig.~\ref{fig:compare_Vcb}. Using the inclusive determination of
$|V_{ub}|$ would imply an even larger value for $|V_{cb}|$.

\section{Ongoing work}
\label{sec:ongoing}

At the Lattice 2016 symposium, there were several talks reporting on
calculations underway.  The HPQCD collaboration are completing an
analysis of the $B_s \to D_s$ form factors \cite{Monahan:2016qxu}, on
the MILC AsqTad ensembles.  In the future, HPQCD and Fermilab/MILC
plan to extend their study of the $B_{(s)} \to D_{(s)}$ form factors
on the MILC HISQ ensembles, reducing discretization,
quark mass extrapolation, and other uncertainties.

HPQCD are in the final stages of calculating the $B\to D^*$ form factor
$h_{A_1}(w)$ at zero recoil, $w=1$ \cite{Harrison:2016gup}.  This calculation
has been done using the MILC HISQ lattices, so it will be the first new result
statistically independent from the Fermilab/MILC calculations, providing an
important check of the discrepancy shown in Fig.~\ref{fig:compare_Vcb}.

By using two formulations for the $b$ quark, NRQCD and heavy HISQ, to compute
the $B_c \to \eta_c$ and $B_c \to J/\psi$ form factors, the HPQCD
collaboration have a nonperturbative means to determine the normalization
of the NRQCD currents \cite{Colquhoun:2016osw}.  This provides an opportunity
to quantify and possibly reduce a dominant uncertainty in NRQCD calculations.

The RBC-UKQCD collaboration presented preliminary results for the $B_s
\to D_s$ form factors using domain wall fermions for light and charm
quarks plus the RHQ variant of the Fermilab action for the bottom
quark \cite{Flynn:2016vej}.  They plan to extend their work to include
all $B_{(s)}\to D_{(s)}^{(*)}$ form factors. These will be useful since
the  methods used are different from Fermilab/MILC and HPQCD.

The calculation of the $B \to D^*$ form factors away from zero recoil is
something several groups are pursuing.  The main motivation is to have
better control over fits to the shape of the differential branching fraction.
To date, $|V_{cb}|$ has been obtained by fitting the experimental data and
extrapolating to zero recoil, at which point the lattice calculation described
above gives the normalization.  Recent experience extracting $|V_{ub}|$ and
$|V_{cb}|$ has underscored the efficacy of using \textit{both}
lattice and experimental information over a range of lepton invariant mass
$q^2$.  The challenges extending this method to the vector meson final state
come from having to determine more form factors, and to overcome larger
statistical errors in the lattice calculations.  It is likely that progress
will first come for the $B_s\to D_s^*$ form factors, making a measurement
of the $B_s\to D_s^*\ell\nu$ differential decay rate very desirable.
Eventually these calculations will also refine the Standard Model prediction
for the $\tau/\ell$ final state ratio $R(D^*)$.

\section*{Acknowledgments}

I am grateful for informal updates from A.\ Kronfeld and O.\ Witzel, 
and for discussions with my collaborators, C.\ Davies, J.\ Harrison,
A.\ Lytle, C.\ Monahan, and J.\ Shigemitsu.
I am supported in part by STFC grant ST/L000385/1.


\bibliographystyle{h-physrev5}
\bibliography{mbw}

\begin{thebibliography}{10}

\bibitem{Amhis:2012bh}
Y.~Amhis {\em et~al.},
\newblock (2012), arXiv:1207.1158.

\bibitem{Bailey:2014tva}
J.~A. Bailey {\em et~al.},
\newblock Phys. Rev. D {\bf 89}, 114504 (2014), arXiv:1403.0635.

\bibitem{Abdesselam:2017kjf}
A.~Abdesselam {\em et~al.},
\newblock (2017), arXiv:1702.01521.

\bibitem{Bevan:2014iga}
A.~J. Bevan {\em et~al.},
\newblock Eur. Phys. J. C {\bf 74}, 3026 (2014), arXiv:1406.6311.

\bibitem{Alberti:2014yda}
A.~Alberti {\em et~al.},
\newblock Phys. Rev. Lett. {\bf 114}, 061802 (2015), arXiv:1411.6560.

\bibitem{Bailey:2015rga}
J.~A. Bailey {\em et~al.},
\newblock Phys. Rev. D {\bf 92}, 034506 (2015), arXiv:1503.07237.

\bibitem{Na:2015kha}
H.~Na {\em et~al.},
\newblock Phys. Rev. D {\bf 92}, 054510 (2015), arXiv:1505.03925,
\newblock [Erratum: Phys. Rev. D \textbf{93}, 119906 (2016)].

\bibitem{Aoki:2016frl}
S.~Aoki {\em et~al.},
\newblock Eur. Phys. J. C {\bf 77}, 112 (2017), arXiv:1607.00299.

\bibitem{Glattauer:2015teq}
R.~Glattauer {\em et~al.},
\newblock Phys. Rev. D {\bf 93}, 032006 (2016), arXiv:1510.03657.

\bibitem{Bailey:2012jg}
J.~A. Bailey {\em et~al.},
\newblock Phys. Rev. Lett. {\bf 109}, 071802 (2012), arXiv:1206.4992.

\bibitem{Bailey:2012rr}
J.~A. Bailey {\em et~al.},
\newblock Phys. Rev. D {\bf 85}, 114502 (2012), arXiv:1202.6346,
\newblock [Erratum: Phys. Rev. D \textbf{86}, 039904 (2012)].

\bibitem{Atoui:2013zza}
M.~Atoui {\em et~al.},
\newblock Eur. Phys. J. C {\bf 74}, 2861 (2014), arXiv:1310.5238.

\bibitem{Aaij:2015bfa}
R.~Aaij {\em et~al.},
\newblock Nature Phys. {\bf 11}, 743 (2015), arXiv:1504.01568.

\bibitem{Detmold:2015aaa}
W.~Detmold, C.~Lehner, and S.~Meinel,
\newblock Phys. Rev. D {\bf 92}, 034503 (2015), arXiv:1503.01421.

\bibitem{Monahan:2016qxu}
C.~J. Monahan {\em et~al.},
\newblock Proc. Sci. {\bf LATTICE2016}, 298 (2016), arXiv:1611.09667.

\bibitem{Harrison:2016gup}
J.~Harrison, C.~Davies, and M.~Wingate,
\newblock Proc. Sci. {\bf LATTICE2016}, 287 (2016), arXiv:1612.06716.

\bibitem{Colquhoun:2016osw}
B.~Colquhoun {\em et~al.},
\newblock Proc. Sci. {\bf LATTICE2016}, 281 (2016), arXiv:1611.01987.

\bibitem{Flynn:2016vej}
J.~Flynn {\em et~al.},
\newblock Proc. Sci. {\bf LATTICE2016}, 296 (2016), arXiv:1612.05112.

\end{thebibliography}


\end{document}